\documentclass[aps,prb,twocolumn,groupedaddress]{revtex4-1}
\usepackage[dvips]{graphicx}
\usepackage{amsmath}
\usepackage{color}
\usepackage{bm}

\begin{document}


\title{Real-Time Dynamics of a Spin Chain with Dzyaloshinskii-Moriya Interactions: Spiral Formation and Quantum Spin Oscillations}


\author{E. Solano-Carrillo}
\author{R. Franco}
\author{J. Silva-Valencia}
\affiliation{Departamento de F\'{\i}sica, Universidad Nacional de Colombia, A.A. 5997, Bogot\'a-Colombia.}


\date{\today}

\begin{abstract}
We studied the non-equilibrium short-time dynamics of a spin-$1/2$ chain with Dzyaloshinskii-Moriya interactions after a sudden quench by a transverse 
field. We found that inhomogeneous spin-spirals with opposite chiralities propagate from the edges towards the center of the chain. This propagation is 
accompanied by quantum spin oscillations which decay asymptotically with time. A theoretical description of this phenomenon is given to a good accuracy 
with the help of numerical calculations with the adaptive time-dependent density matrix renormalization group algorithm.
\end{abstract}


\maketitle

\section{Introduction}
In recent years, many attempts have been made to understand the non-equilibrium dynamics of closed interacting quantum systems \cite{Polkovnikov}. These 
efforts are mainly motivated by the high degree of isolation and control of microscopic parameters which can be achieved experimentally in systems such 
as ultracold atoms trapped in optical lattices\cite{Bloch}, which are currently being used to ``engineer'' quantum spin systems with desired 
properties\cite{Duan,Trotzky,Simon} with the aim of simulating the behavior of emblematic systems in quantum statistical physics and discovering 
potential technological applications. Among the most studied problems which have been addressed, the dynamics of low-dimensional systems following a 
\emph{quantum quench} has received special attention. This is one of the simplest ways to drive a system far from equilibrium. It consists of preparing 
the system in the ground state of an initial Hamiltonian and then suddenly (or slowly) changing the interactions at a given time, by varying an 
experimentally tunable parameter, with the result of letting the system evolve unitarily under a final Hamiltonian. As an example, one might mention the 
quenching dynamics by a transverse magnetic field (tunable parameter) in systems like the XY chain\cite{Mukherjee,Mukherjee2,Sengupta,GuoXY} or the 
quantum Ising chain\cite{Rossini,Calabrese,Calabrese2,Das,Li}, which can be simulated in ultracold atomic systems\cite{Simon}.\par
Quenching by a transverse field is a typical protocol for studying non-equilibrium dynamics, since apart from being easy to implement in real situations, it 
allows the possibility of taking the system (in some cases) through a quantum phase transition, involving \emph{macroscopic} changes in the many-body 
state at the final point, as in the quantum Ising chain\cite{Simon}. In more general quenching scenarios, the universal properties of the dynamics near 
quantum critical points as well as in generic gapped and gapless systems have been investigated thoroughly\cite{Polkovnikov}; in some cases it is 
possible to make universal predictions of scaling laws for relevant physical quantities. Also, the question of whether the system relaxes to a 
quasi-stationary state in the long-time limit after the quench and, if it does, how to characterize its physical properties, has been addressed with a 
great interest. In particular, the concept of \emph{thermalization} of a quantum many-body system following a quench has acquired importance; it has 
been conjectured that the asymptotic state can be described by a generalized Gibbs ensemble\cite{Rigol,Rigol2} if the system is integrable, and by some 
effective thermal distribution in the case of a generic system. Another main line of research, which we follow in this paper, is to study the 
characteristics of the time evolution just after the quench (short-time dynamics). This is of special interest, for example, in quantum computation 
scenarios\cite{Bennett} in which gate operations require a precise knowledge of the immediate dynamical response of the system. Experimental examples 
following this line of research are the oscillations\cite{Greiner} or the dephasing\cite{Hofferberth} of the matter wave field of a Bose-Einstein 
condensate, features which are also observed in the antiferromagnetic order parameter of a system of ultracold atoms in an optical lattice with 
superexchange interactions\cite{Trotzky}.\par
There are many spin systems which serve as prototypes for studying the short-time non-equilibrium dynamics following a quantum quench. In this paper we 
study a spin system with exchange-relativistic interactions which, according to our knowledge, has not been investigated so far in this context. 
These kind of interactions arise when Anderson's superexchange mechanism is extended to include spin-orbit coupling. As is well known, the 
spin-orbit interaction (which couples the spin to the crystal lattice, giving rise to easy and hard magnetization axes) is gaining considerable 
attention due to the possibility of manipulating spins in solid state systems solely by electric fields \cite{Awschalom} and/or perhaps by applied 
local strain \cite{Dreher,Lovett}. In low-dimensional systems, the lack of structural inversion symmetry which is induced by surfaces and interfaces, 
combined with the spin-orbit coupling leads to anisotropic \emph{antisymmetric} superexchange or Dzyaloshinskii-Moriya interaction (DMI) 
\cite{Dzyaloshinsky,Moriya}, which is responsible for the presence of weak ferromagnetism in a variety of antiferromagnetic compounds\cite{Moriya}. 
One of the most striking consequences of this spin interaction is the appearance of homochiral spin spiral magnetic structures, as have been observed 
recently along fixed crystallographic directions (easy-axis directions in some cases) in ferromagnetic and antiferromagnetic ultrathin films using 
spin-polarized scanning tunneling microscopy \cite{Bode,Ferriani,Meckler}. These one-dimensional spin structures have been postulated to play a crucial 
role in spintronic devices, since spin-polarized currents flowing through them will exert a spin-torque on the chiral magnetic structure, causing a 
variety of controllable excitations \cite{Bode}. They are also the most likely candidates to host ferroelectricity, according to recent experimental 
results\cite{Kimura,Lawes,Kenzelmann}. In the context of ultracold atoms in optical lattices, it remains to be discovered whether the DMI can be 
artificially implemented and controlled, as can be done with the \emph{symmetric} superexchange interaction \cite{Trotzky}. Due to the great 
interest in the consequences of this antisymmetric interaction, this is expected to be achieved soon, since the spin-orbit coupling has already been 
demonstrated recently in these systems of atoms\cite{Lin}.\par
Motivated by the above-mentioned potential applications as well as by the search for fundamental knowledge, in this paper we study the short-time non-equilibrium 
dynamics in the simplest system with DMI: a spin-$1/2$ chain, after a quantum quench is carried out by means of a transverse field. To be more specific, 
we consider a quantum quench which consists of preparing the initial state $|\psi_0\rangle=\mid\rightarrow\rightarrow\dots\rightarrow\rangle$ by 
applying a strong magnetic field along the $x$-direction, and then turning it off abruptly at $t=0$. The subsequent unitary evolution, 
$|\psi_t\rangle=\exp(-i\mathcal{H}t)|\psi_0\rangle$, is governed by the final Hamiltonian (we set $\hbar=1$ throughout this paper)
\begin{equation}\label{HDJ}
\begin{split}
 \mathcal{H}=\sum_j\,[\,&D\,(S_j^xS_{j+1}^y-S_j^yS_{j+1}^x)\\
  &+J\,(S_{j}^xS_{j+1}^x+S_{j}^yS_{j+1}^y+\Delta\,S_{j}^zS_{j+1}^z)\,],
\end{split}
\end{equation}
where the usual DMI in the $z$-direction is taken together with the Heisenberg-type symmetric exchange coupling (EC) with anisotropy $\Delta$ (also 
known as the XXZ exchange interaction). As an example of a real quasi-one-dimensional compound described by this Hamiltonian we mention KCuF$_3$, which 
is an antiferromagnetic spin-$1/2$ system with DMI and almost isotropic EC \cite{Yamada}. The critical behavior of this system will be briefly 
discussed in section \ref{secDJ}. Meanwhile, we simply mention that the DMI makes the phase diagram rich, with two critical points (as in the XXZ model) 
dependent on $D$, and restores the otherwise spoiled ground-state entanglement due to anisotropy (which tends to align the spins), by means of enhanced 
quantum fluctuations\cite{Kargarian2}. With respect to some dynamical and thermal features, this model has attracted attention in the quantum information 
community; for instance, the striking phenomenon of finite-time disentanglement \cite{Yu} has been predicted for the two-qubit system prepared in the
 Werner state \cite{Fang}, and teleportation of states has been studied using the model in thermal equilibrium as a quantum channel \cite{Guo}.\par
In most of the numerical calculations in this paper, we use the adaptive time-dependent density matrix renormalization group algorithm (t-DMRG) 
\cite{White,Daley}, which has proved to be a powerful method for studying the dynamics of one-dimensional quantum many-body systems with short-ranged 
interactions. As a general result, we show that the DMI gives rise to inhomogeneous spin-spirals with opposite chiralities, which propagate from the 
edges towards the center of the chain until they meet each other, and then complex interference patterns arise. The propagation is accompanied by 
quantum spin oscillations that decay asymptotically with time. It is the aim of this work to theoretically describe these oscillations, as well as the 
nature of the propagation of the spin-spirals. The paper is organized as follows: in section \ref{secD} we study in detail the role played only by the 
DMI in the formation of these spirals. For this, we use results from the two-spin quantum dynamics and many-body classical dynamics to interpret the 
observations from the quantum model. In section \ref{secDJ}, we add the exchange interactions and investigate how the results from section \ref{secD} 
are modified. In section \ref{secLT}, we make some comments about the long-time dynamics in our system; and finally in section \ref{secC}, we give the 
conclusions. \par
\section{Effects of the DMI}\label{secD}
Although the DMI is usually expected to be negligible (due to its relativistic nature) compared to the EC, there are instances \cite{Bode,Ferriani} 
where it is strong enough ($D/J\sim0.25$) to create remarkable spin-spiral structures. In this section, we take the limit $J=0$ in order to investigate 
the effects on the dynamics due only to the DMI. Then Eq. \eqref{HDJ} is reduced to
\begin{equation}\label{HDMI}
 \mathcal{H}^{\textrm{DM}}=D\sum_j\,(S_j^xS_{j+1}^y-S_j^yS_{j+1}^x).
\end{equation}
This model is exactly solvable, since it can be mapped onto the XX model: 
$\mathcal{H}^{\textrm{XX}}=D\sum_j\,(\tilde{S}_j^x\tilde{S}_{j+1}^x+\tilde{S}_j^y\tilde{S}_{j+1}^y)$ (which is well known to be equivalent to a set of 
free spinless fermions) by rotating the basis at each site $j$ by an angle $j\frac{\pi}{2}$ around the $z$-axis. Moreover, its eigenstates are 
non-equilibrium or spin-current-carrying stationary states of the XX model at zero temperature, since the macroscopic spin-current $J^M=\sum_jJ_j^M$ 
(quantized along the direction of $D$) defined through the continuity equation
\begin{equation}
 \dot{S}_j^z=i\left[\mathcal{H}^{\textrm{XX}},S_j^z\right]=J_j^M-J_{j-1}^M,
\end{equation}
can be found\cite{Antal} to give $J^M=\mathcal{H}^{\textrm{DM}}$ and, given that this spin-current is a conserved quantity in the XX model, it follows 
that $\left[\mathcal{H}^{\textrm{XX}},\mathcal{H}^{\textrm{DM}}\right]=0$.
The time evolution resulting from several initial states in the XX model such as the domain-wall state at the center of the 
chain\cite{Antal,Hunyadi,Gobert}, the N\'eel state and the spin-density wave state \cite{Barmettler} has already been solved analytically, yielding  
a variety of phenomena such as quantum fronts in which quantized steps of spin flips are transported and oscillatory dynamics of the order parameter. 
Our initial spin-polarized state in the $x$-direction $|\psi_0\rangle\propto(\mid\uparrow\rangle+
\hspace{-1mm}\mid\downarrow\rangle)\otimes(\mid\uparrow\rangle+\hspace{-1mm}\mid\downarrow\rangle)\cdots\otimes(\mid\uparrow\rangle+
\hspace{-1mm}\mid\downarrow\rangle)$ contains eigenstates of the XX model such as the fully polarized states 
$\mid\uparrow\uparrow\cdots\uparrow\rangle$ and $\mid\downarrow\downarrow\cdots\downarrow\rangle$, and also many excitations with a large number of 
domain walls (the N\'eel state being one of those); thus it is a highly non-equilibrium state at zero temperature. When this initial state is evolved 
under the XX Hamiltonian, there is no signature of dynamical formation of spin-spirals. Otherwise, when it is evolved under the DM Hamiltonian, the 
trajectory in the Hilbert space of the resulting state is different, and the formation of the spin-spirals is observed. Since there are no driving 
fields to excite this wave phenomenon, the open boundary conditions in our finite system play a crucial role in the establishment of these 
spin-currents. In particular, since $\langle\psi_0|J^M|\psi_0\rangle=0$, we expect to encounter a positive spin-current propagating to the right from 
one boundary, and a negative spin-current propagating to the left from the other boundary. Although in principle the real-time evolution can be 
investigated analytically in this case, it is a highly non-trivial problem; therefore we use the t-DMRG algorithm to inquire further into this.\par
\begin{figure}
 \centering
\includegraphics[scale=0.55]{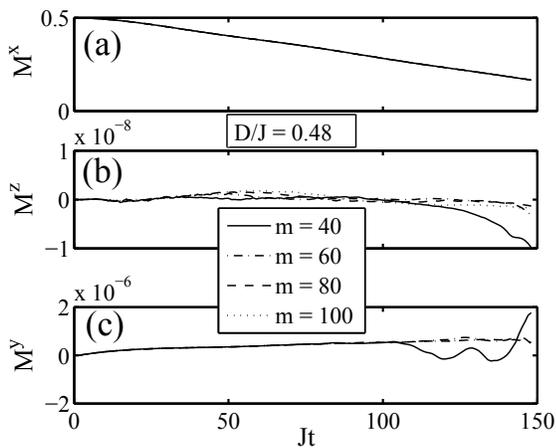}
\caption{Time evolution of the components of magnetization for several values of the t-DMRG truncation parameter $m$ in the arbitrary case with DMI and 
EC for which $D/J=0.48$. (a) $x$-component. (b) $z$-component. (c) $y$-component.\label{Mvsm}}
\end{figure}
\subsection{Many-body quantum dynamics}
We have used the t-DMRG algorithm for a chain with $N=200$ sites, with a Trotter time step $\delta t=0.05$, and keeping $m=40-100$ density matrix 
eigenstates to represent the truncated sector of the evolved wave function. To test the convenience of this choice of simulation parameters, we show in 
Fig. \ref{Mvsm} the time evolution of the magnetization $M^{\alpha}=(1/N)\sum_j\langle S_j^{\alpha}\rangle$ ($\alpha=x,y,z$) for several values of $m$, 
in an arbitrary case with $J\neq0$ in which strong long-ranged correlations are expected to lower the accuracy of the algorithm compared to the case 
$J=0$. As can be seen, for values $m\geq60$, the $m$-dependent error in the chosen time window is of order $10^{-7}$ in the worst case, which is a 
satisfactory error for DMRG users. In the following calculations, we used $m=60$ (unless otherwise stated) with the aim of saving computer resources and 
speeding them up. Note that when the value of $J$ is not shown in the figures, it is assumed to be zero.\par
\begin{figure}
 \centering
\includegraphics[scale=0.55]{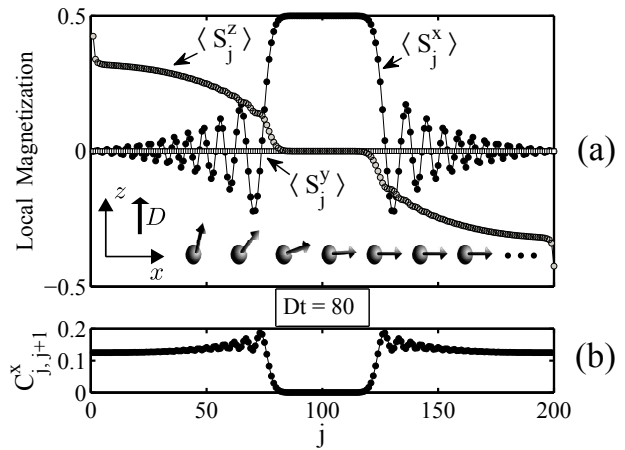}
\caption{Snapshot of the local magnetization and nearest-neighbor spin-spin correlations in the $x$-direction at the arbitrary time $Dt=80$. (a) As can 
be seen, inhomogeneous spin-spirals with opposite chiralities propagate, in the $xz$-plane, from the edges towards the center of the chain, accompanied 
by spin oscillations. A classical picture of the left-rotating spiral arising from the left end is drawn when $Dt=2$ (only the directions of spins are 
taken into account). (b) The spirals propagate with the same velocity with which nearest-neighbor correlations move.\label{Szi-FE}}
\end{figure}

Returning to the case of relevance for this section ($J=0$), we show in Fig. \ref{Szi-FE} a ``snapshot'' of the local magnetization along the chain as 
well as the nearest-neighbor spin-spin correlations $C_{j,j+1}^x=\langle S_j^xS_{j+1}^x\rangle-\langle S_j^x\rangle\langle S_{j+1}^x\rangle$ in the 
direction of the polarization of the initial state, at the arbitrary time $Dt=80$. As can be seen, inhomogeneous spin-spirals with opposite chiralities 
propagate (in the $xz$-plane) from the edges towards the center of the chain, as expected from the spin-current considerations given above. This 
propagation is accompanied by spin oscillations around the $z$-axis. A sketch of the left-rotating spiral propagating from the left end of the chain is 
drawn when $Dt=2$. We verified that the chirality (sense of rotation) of the spin-spirals reverses when the sign of $D$ is reversed, which is an 
inherent property of the DMI \cite{Meckler}. We also followed the wave fronts, and found a constant velocity of propagation $v=1$ 
(in units of $aD$, $a$ being the lattice constant). As can be seen in Fig. \ref{Szi-FE}(b), this is the same velocity with which nearest-neighbor 
correlations propagate from the edges, in accordance with the light-cone propagation which is expected for spin models with sufficiently local 
interactions \cite{Lieb,CalaCardy}.\par
It is also interesting to investigate the time evolution of long-ranged correlations, which we calculate as is usually done in the DMRG 
framework\cite{White2,Manmana} $C_r^x=\langle S_i^xS_j^x\rangle-\langle S_i^x\rangle\langle S_j^x\rangle$, where $i=N/2+1-r$ and $j=N/2+r$ (when $r$ is 
increased, $i$ and $j$ move outwards symmetrically from the center towards the edges of the chain). We found that long-ranged correlations are quite 
small in this case: they are always below O($10^{-3}$) during the spiral propagation, compared to the nearest-neighbor ones which, as may be seen in 
Fig. \ref{Szi-FE}(b), are O($10^{-1}$); this is why they are not shown. We shall see later that when $J\neq0$, strong long-ranged correlations 
develop throughout the system, making the dynamics more intricate. Meanwhile, let us first try to understand the nature of the results obtained so far.
\subsection{Two-spin quantum dynamics}
\begin{figure}
 \centering
\includegraphics[scale=0.55]{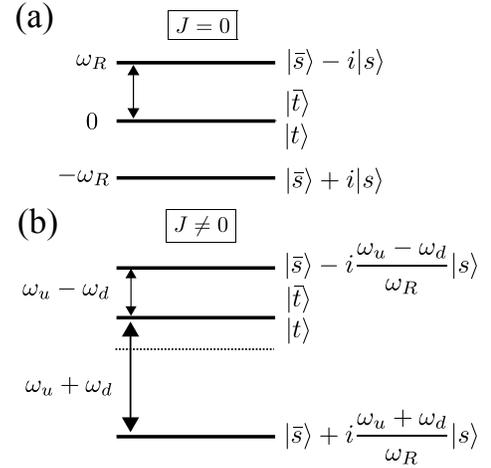}
\caption{Energy levels and stationary states of the two-spin model with DMI. (a) In the absence of EC, the first excited state is two-fold degenerate. 
(b) In the presence of EC, the first excited state remains degenerate, but the energy gaps are modified.\label{Elev}}
\end{figure}

We begin by considering the quantum dynamics of a two-spin system, which should contain relevant information about the behavior of a pair of 
nearest-neighbor sites in the many-body system. In order to simplify the description of states, we rotate the basis temporarily by changing 
$x\leftrightarrow z$ in Eq.  \eqref{HDJ}, which leads to a DMI in the $x$-direction and a fully polarized initial state in the $z$-direction. Note that 
this operation does not change expected values of observables. It is also convenient to introduce the Bell states in this new representation, defined 
as $|t/s\rangle=(\mid\uparrow\downarrow\rangle\pm\mid\downarrow\uparrow\rangle)/\sqrt{2}$ and 
$|\bar{t}/\bar{s}\rangle=(\mid\uparrow\uparrow\rangle\pm\mid\downarrow\downarrow\rangle)/\sqrt{2}$. Then we show in Fig. \ref{Elev}(a) the energy 
levels and the stationary states of the two-spin model, with $\omega_R=D/2$. Energetically speaking, this is equivalent to a two-level magnetic system 
with a spinless intermediate state. In the present case ($J=0$), when the initial state $|\psi_0\rangle=\mid\uparrow\uparrow\rangle$ is prepared, the 
evolved two-spin state reads
\begin{equation}\label{Psi2}
 |\psi_t\rangle_{2s}=\dfrac{1}{\sqrt{2}}\left[\,|\bar{t}\,\rangle+\cos(\omega_Rt)|\bar{s}\rangle-\sin(\omega_Rt)|s\rangle\,\right],
\end{equation}
which shows Rabi oscillations between the states $\mid\uparrow\uparrow\rangle$ and $\mid\downarrow\downarrow\rangle$, i.e., the $z$-components of local 
magnetization for the spin pair oscillate in phase $\langle S_1^z\rangle=\langle S_2^z\rangle=\frac{1}{2}\cos(\omega_Rt)$ with the Rabi frequency 
$\omega_R$. Remarkably, the most probable state to find after an energy measurement of the spin pair (during a cycle) is the spinless intermediate Bell 
state $|\bar{t}\,\rangle$, so, at any time, spin measurements in the direction perpendicular to $D$ will disperse around zero, as indeed is seen  
in Fig. \ref{Szi-FE}(a).\par
Returning to the original representation, we have a two-spin system with DMI in the $z$-direction and initially prepared in the state 
$|\psi_0\rangle=\mid\rightarrow\rightarrow\rangle$. The time evolution shows Rabi oscillations between the states $\mid\rightarrow\rightarrow\rangle$ 
and $\mid\leftarrow\leftarrow\rangle$ with $\langle S_1^x\rangle=\langle S_2^x\rangle=\frac{1}{2}\cos(\omega_Rt)$. A natural question then arises: how 
can we generalize these results to many sites? It is reasonable to assume that, at least for times $Dt<1$, the evolved state is close to
\begin{equation}\label{Dimers}
|\psi_t\rangle\stackrel{?}{\sim}\;\, |\psi_t^{+}\rangle_{2s}\,
\otimes\mid\rightarrow\rightarrow\rangle\otimes\cdots\otimes\mid\rightarrow\rightarrow\rangle\otimes|\psi_t^{-}\rangle_{2s},
\end{equation}
where $|\psi_t^{+}\rangle_{2s}$ is the state in Eq. \eqref{Psi2} written in the original representation and $|\psi_t^{-}\rangle_{2s}$ is the 
corresponding state obtained when changing $D\leftrightarrow-D$, in order to account for the opposite chirality of the spiral arising from the right 
end of the chain (negative spin-current). This assumption is motivated by the fact that for $Dt<1$, spin-spin correlations have not crossed the 
boundaries of the two-site blocks at the ends of the chain, so the state can be thought of as factorized into ``non-interacting dimers'': 
$\boxed{\textrm{\textbullet}\hspace{3mm}\textrm{\textbullet}}\hspace{1.5mm}\boxed{\textrm{\textbullet}\hspace{3mm}\textrm{\textbullet}}\;\;\cdots\;\;
\boxed{\textrm{\textbullet}\hspace{3mm}\textrm{\textbullet}}\hspace{1.5mm}\boxed{\textrm{\textbullet}\hspace{3mm}\textrm{\textbullet}}\,$, i.e., a 
product of two-spin solutions. The problem with this assumption is that it does not take into account the spiral propagation; i.e., since the 
perturbations originated at the chain boundaries propagate with a finite velocity, it is clear that the spins within these dimers must oscillate with 
different phases, so the state in Eq. \eqref{Dimers} should be modified to accomplish this. \par
\begin{figure}
 \centering
\includegraphics[scale=0.55]{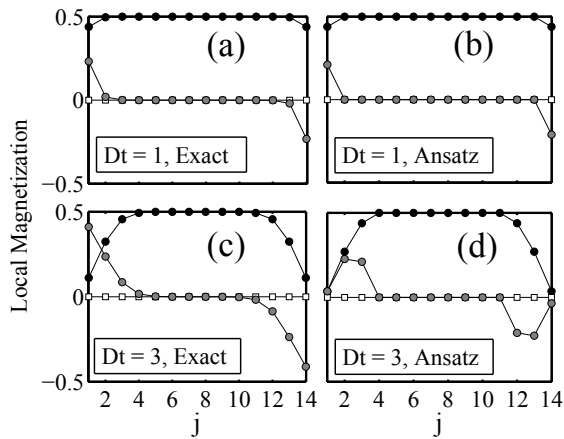}
\caption{Snapshots of the local magnetization for a system with 14 sites, calculated exactly: (a), (c); and with the ansatz in Eq. \eqref{rhot}: (b), 
(d), for two different times. The black-filled circles are values of $\langle S_j^x\rangle$, the gray-filled ones are those of $\langle S_j^z\rangle$, 
and the white-filled squares are values of $\langle S_j^y\rangle$, as in Fig. \ref{Szi-FE}(a). Solid lines are guides for the eyes.\label{Szi-ex}}
\end{figure}
\begin{figure}
 \centering
\includegraphics[scale=0.55]{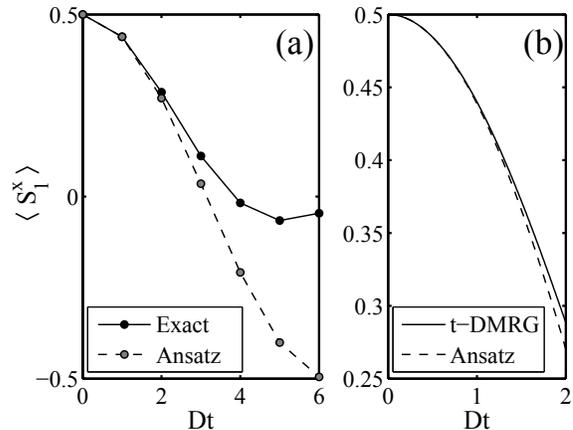}
\caption{Time evolution of the $x$-component of local magnetization for the first spin in a system with 14 sites. (a) Comparison between exact results 
with those obtained from the ansatz in Eq. \eqref{rhot}; solid and dashed lines are guides for the eyes. (b) Finer curve obtained with t-DMRG for times 
$Dt\leq2$.\label{Sz1x-ex}}
\end{figure}

A convenient way to account for the spiral formation is to decouple the two-spin dynamics by introducing the reduced density matrix 
$\rho_{t}^{\pm}=\textrm{Tr}_2|\psi_t^{\pm}\rangle_{2s}\langle\psi_t^{\pm}|$ for the spins that have been reached by the perturbations, and 
$\rho_{0}=\mid\rightarrow\rangle\langle\rightarrow\mid$ for those that have not. Then, an easy calculation leads to 
$\rho_{t}^{\pm}=\frac{1}{2}\mathcal{I}+\cos(\omega_Rt)S^x\pm\frac{1}{2}\sin(Dt)S^z$ and $\rho_{0}=\frac{1}{2}\mathcal{I}+S^x$, where $\mathcal{I}$ is 
the $2\times2$ identity matrix. Note that this decoupling of the two-spin dynamics preserves the relation 
$\langle S^x\rangle=\textrm{Tr}(\rho_{t}^{\pm}S^x)=\frac{1}{2}\cos(\omega_Rt)$. With this, we expect that the evolved state will be close to
\begin{equation}\label{rhot}
 \rho_t\sim \bigotimes_{j=1}^{N/2}\varLambda_j^{+}(t)\hspace{-2mm}\bigotimes_{j=N/2+1}^{N}\hspace{-2mm}\varLambda_j^{-}(t),
\end{equation}
where $\varLambda_j^{\pm}(t)=[1-H(\tau_j)]\rho_0+H(\tau_j)\rho_{\tau_j}^{\pm}$ selects the approximate state of the site $j$ at time $t$, with $H$ 
being the Heaviside step function and $\tau_j=t-(j-1)/D$ ($\tau_j=0$ the moment at which the perturbation hits the site $j$, measured from the chain 
boundaries). We tested the validity of the ansatz in Eq. \eqref{rhot} by calculating the exact time evolution of local magnetization for a system 
with 14 sites, as shown in Fig. \ref{Szi-ex}. As can be seen, the ansatz works very well for times $Dt\le1$ as expected from the arguments motivating 
Eq. \eqref{Dimers}. A closer look at the single-site dynamics is shown in Fig. \ref{Sz1x-ex}, up to the time just before the propagating spirals 
interfere with each other at the center. Fig. \ref{Sz1x-ex}(b) shows the comparison between the $x$-component of the local magnetization calculated 
with the ansatz and with t-DMRG using a Trotter time step $\delta t=0.01$ and $m=100$ (it would take too much time to sweep the exact curve). 
As is evident, the ansatz gives satisfactory results even for times $Dt\le2$. For larger times, the assumption leading to Eq. \eqref{Dimers} is no 
longer valid, since the perturbations have reached the third sites (measured from the chain boundaries), and three-body effects then arise. 
In general, as nearest-neighbor spin-spin correlations propagate towards the center, many-body effects become important and the single-site dynamics 
deviates remarkably from what is expected from Eq. \eqref{rhot}. As we shall discuss next in more detail, the most remarkable many-body effect is the 
onset of an ``emergent'' Rabi frequency of oscillations $\omega_R^{\star}$ in the single-site dynamics; these oscillations decay with time as can be 
anticipated from Fig. \ref{Sz1x-ex}(a).
\subsection{Many-body classical dynamics}
\begin{figure}
 \centering
\includegraphics[scale=0.55]{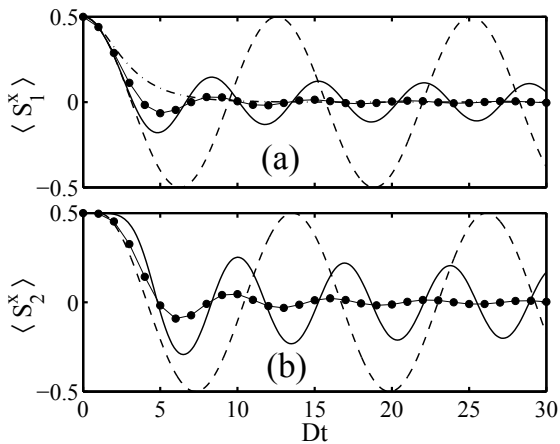}
\caption{Time evolution of the $x$-component of local magnetization for (a) the first spin and (b) the second spin along the chain. The black-filled 
circles are the results from t-DMRG. The dash-dotted line in (a) is the result of the classical two-spin model described by Eq. \eqref{tl}. The solid 
line is the numerical solution from the many-body classical dynamics governed by Eq. \eqref{dteta}, and the dashed line is the prediction from the 
ansatz in Eq. \eqref{rhot}.\label{cmpSz-ed}}
\end{figure}

The simplest method to investigate many-body effects in our system is by appealing to classical considerations. At the limit of commuting dynamical 
variables, the Heisenberg equations of motion for the spin components turn into the Landau-Lifshitz equations (without damping) for the dynamics of 
classical spin vectors $\bm{S}_j$ precessing around effective fields $\bm{B}_j$:
\begin{equation}
 \dfrac{d\bm{S}_j}{dt}=-\bm{S}_j\times\bm{B}_j, \;\;\textrm{with}\;\;\bm{B}_j=\dfrac{\partial\mathcal{H}^{\textrm{DM}}}{\partial\bm{S}_j}.
\end{equation}
In polar coordinates $S_j^x=\frac{1}{2}\cos\theta_j$, where $\theta_j$ is the propagating spiral profile, and if we assume $S_j^y=0$, as is evident 
from Figs. \ref{Szi-FE}(a) and \ref{Szi-ex}, we obtain the system of differential equations
\begin{equation}\label{dteta}
 \dot{\theta}_j=\omega_R(\cos\theta_{j+1}-\cos\theta_{j-1}).
\end{equation}
Let us consider the classical two-spin model first. Just after $t=0$, the first two spins are almost collinear and thus $\theta_1\approx\theta_2$. With 
this approximation, we have $\dot{\theta}_1=\omega_R\cos\theta_1$ and $\dot{\theta}_2=-\omega_R\cos\theta_2$, with the solution
\begin{equation}\label{tl}
 2S_1^x(t)=2S_2^x(t)=\textrm{sech}(\omega_Rt).
\end{equation}
This is plotted in Fig. \ref{cmpSz-ed}(a) with a dash-dotted line. It suggests that the classical spin pair (decoupled from the many-body system) moves 
in phase, as is the case in the quantum two-spin solution. However, as may be seen from the figure, the classical dynamics of the spin pair do not 
display any kind of oscillation. As we shall see next, this is not the case in the many-body classical system.\par
The first step for solving the many-body classical dynamics is to numerically solve the system of differential equations in Eq. \eqref{dteta}. 
From this system we observe that, since $\dot{\theta}_1=\omega_R\cos\theta_2$ and $\dot{\theta}_N=-\,\omega_R\cos\theta_{N-1}$, the origin of 
spin-spirals with opposite chiralities at the chain boundaries is a straightforward consequence of the classical open boundary conditions (see the 
change of signs). The solution of these equations for the first two spins along the chain is shown in Fig. \ref{cmpSz-ed} with thick solid lines. As 
can be seen, the classical solution is able to predict the decaying oscillations over time, although the decay is far slower than the data from the 
quantum model (results from t-DMRG). Also, the observed frequency of the oscillations $\omega_R^{\star}$ is very well reproduced by the classical 
solution. Comparing that with $\omega_R$ obtained from the ansatz in Eq. \eqref{rhot} (dashed lines), which neglects many-body correlations, we have 
$\omega_R^{\star}=2\omega_R$. Thus we conclude that the Rabi oscillations in the quantum two-spin system persist in the many-body system, but with a 
different ``emergent'' frequency.  This kind of behavior has appeared recently\cite{Barmettler} in the context of quantum quenches in the anisotropic 
spin-$1/2$ Heisenberg chain. There, when the initial N\'eel state $\mid\uparrow\downarrow\rangle$ is prepared and subsequently evolved under the 
two-spin XX Hamiltonian, Rabi oscillations (between the states $\mid\uparrow\downarrow\rangle$ and $\mid\downarrow\uparrow\rangle$) are observed in 
the staggered magnetization (antiferromagnetic order parameter) with a frequency $\omega_R=J$. However, when a sudden quench is performed from the 
many-body N\'eel state to the critical phase in the XX model, the order parameter still displays Rabi oscillations, but with the emergent frequency 
$\omega_R^{\star}=2\omega_R$, as in the present case.\par
The classical single-site spin oscillations coming from the many-body numerical solution take place with the same frequency as in the quantum model, 
although a slight dephasing between the two results can be observed in Fig. \ref{cmpSz-ed}. This is because the classical solution is not able to 
reproduce the velocity $v=aD$ with which the spin-spirals propagate (see e.g. the beginning of the corresponding curves in Fig. \ref{cmpSz-ed}(b)). We 
said before that the present quantum model is equivalent to a set of free spinless fermions. In fact, the resulting dispersion relations are the same 
as those of the XX model, but shifted by the wave number $k_a=\frac{\pi}{2}a^{-1}$, since
\begin{equation}
 \mathcal{H}^{\textrm{DM}}=\sum_k\omega_R^{\star}\sin(ka)\,c_k^{\dagger}c_k-\sum_k\omega_Re^{-ia(k-k_a)},
\end{equation}
where $c_k^{\dagger}$ ($c_k$) are creation (anihilation) operators of free spinless fermions with momentum $k$. Within the thermodynamic limit, the second 
term of this expression vanishes, and the group velocity of propagating excitations is given by $v_k=aD\cos(ak)$. Then, the propagating spin-spirals 
can be associated with free-fermion excitations with zero total momentum (zero energy modes) giving $v_0=v=aD$. Hence, since this velocity comes from a 
purely quantum phenomenon, it is reasonable that classical considerations do not reproduce it.\par
In summary, we have learned from this section that the DMI term only, being exactly spin-current (quantized along the direction of $D$), drives the 
formation of spin-spirals which can be described qualitatively in classical terms, but needs input from quantum theory to fully account for the observed 
behavior. In particular, the two-body quantum dynamics are helpful for interpreting the more complex quantum many-body results. In the following section, we 
investigate how these results are modified when isotropic EC is added to the system.
\section{Adding exchange interactions}\label{secDJ}
In order to investigate the effects of EC on the non-equilibrium dynamics studied above, we consider some limits of the anisotropic Heisenberg exchange 
interaction in Eq. \eqref{HDJ}. First note that, by rotating the basis on odd (or even) sites by an angle $\pi$ around the $z$-axis, the 
antiferromagnetic case of the Hamiltonian ($J>0$) can be mapped onto the ferromagnetic case ($J<0$) with opposite sign of the anisotropy $\Delta$. In 
this paper we take $J>0$ as our overall energy scale (measuring time in units of $J^{-1}$), and give a brief overview of the quantum phase diagram as 
the non-dimensional parameters $\Delta$ and $D/J$ are varied. For numerical calculations we set $J=1$. The critical behavior of the (equilibrium) 
ground state of the model in Eq. \eqref{HDJ} is well known\cite{Alcaraz,Zvyagin,Langari2}. For $\Delta\leq-\sqrt{1+(D/J)^2}$, the ground state of the 
system is ferromagnetic. For $-\sqrt{1+(D/J)^2}<\Delta<\sqrt{1+(D/J)^2}$, the energy spectrum is gapless and the ground state does not support any 
kind of long-ranged order and is thus known as a spin fluid. For $\Delta>\sqrt{1+(D/J)^2}$, the ground state is antiferromagnetic (N\'eel ordered). 
In both the spin fluid and N\'eel phases, the helical spin structure in the $xy$-plane is obtained, which is one of the characteristic properties of 
the DMI. At the limit $\Delta\rightarrow\infty$, the Ising model with DMI is realized. The critical properties and entanglement of this model have 
been studied recently\cite{Jafari,Tatiana}, revealing two quantum phases: the antiferromagnetic one and the saturated chiral one, separated by the critical 
point $D/J=1$.
At the limit $\Delta=0$, the XX model with DMI is obtained. Note that by rotating the basis at each site $j$ by an angle $(j-1)\tan^{-1}(D/J)$ around 
the $z$-axis, it can be shown\cite{Derzhko} that the DMI term of the Hamiltonian can be removed, leading to a XX model with a renormalized exchange 
interaction energy $\sqrt{J^2+D^2}$. In the isotropic case ($\Delta=1$), the DMI term can also be removed from the Hamiltonian\cite{Aristov} (in fact, 
this can be done for any value of $\Delta$; see e.g. [\onlinecite{Alcaraz}]) after a similar canonical transformation. The resulting Hamiltonian is 
that of the XXZ Heisenberg model 
\begin{equation}\label{XXZ}
 \mathcal{H}^{\textrm{XXZ}}=\sum_j[\,J^{xx}(\tilde{S}_j^x\tilde{S}_{j+1}^x+\tilde{S}_j^y\tilde{S}_{j+1}^y)+J^z\tilde{S}_j^z\tilde{S}_{j+1}^z\,],
\end{equation}
with planar exchange energy $J^{xx}=\sqrt{J^2+D^2}$, and anisotropy energy $J^{z}=J$. Since $J^z/J^{xx}<1$ for any non-zero value of $D$, the isotropic 
system with DMI will be in a spin-fluid phase\cite{Langari2}, as can be inferred from the discussion above.\par
\begin{figure}
 \centering
\includegraphics[scale=0.55]{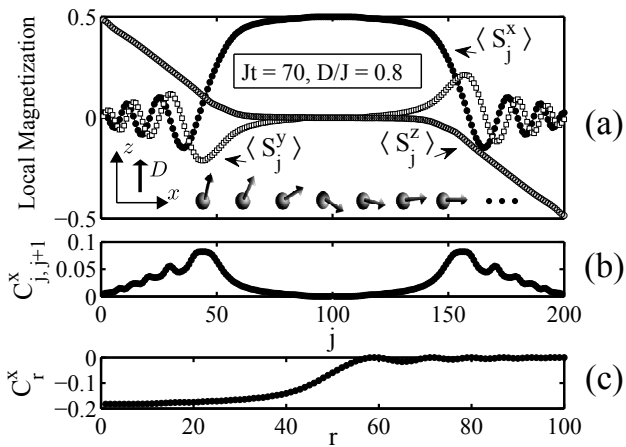}
\caption{(a) Snapshot of the local magnetization at the arbitrary time $Jt=70$, in the model of Eq. \eqref{HDJ} with $\Delta=1$ and $D/J=0.8$. 
A classical picture of the left-rotating spiral arising from the left end of the chain is drawn when $Jt=2$, projected on the $xz$-plane (only the 
directions of spins are taken into account). (b) Nearest-neighbor spin-spin correlations. (c) Spin-spin correlations measured symmetrically from the 
center of the chain.\label{Szj-JD}}
\end{figure}
The relaxation dynamics of the antiferromagnetic order parameter after several quantum quenches in the XXZ model has been investigated in detail 
recently\cite{Barmettler}, revealing a new mode of many-body dynamics in the effectively interacting case of finite anisotropy ($J^z/J^{xx}$ in 
Eq. \eqref{XXZ}): a \emph{exponential} decay of the staggered magnetization with (or without) oscillatory behavior. In a mean-field approximation, a 
rough picture of the dynamics can be obtained by introducing pseudo-spins in $k$-space. Starting from the N\'eel state (equivalent to all pseudo-spins 
pointing in the $x$-direction), the pseudo-spins begin to precess around a ($k,t$)-dependent Zeeman field (static, and in the $z$-direction in the 
effectively non-interacting case of the XX model). Since the field (and hence the precession frequencies) varies continuously from pseudo-spin to 
pseudo-spin over a bandwidth, they gradually \emph{dephase}, causing the oscillations to decay with time. A similar quasi-classical picture (in real 
space) has been described recently in the study of coherent transport of spin currents in the spin-$1/2$ Heisenberg chain in transversed magnetic 
fields\cite{Steinigeweg}. Due to the connection with a real material like KCuF$_3$, in the following discussion, we restrict ourselves to the isotropic case 
($\Delta=1$) of the Hamiltonian in Eq. \eqref{HDJ} (equivalent to Eq. \eqref{XXZ}), focusing on the description of the frequency of single-site spin 
oscillations by means of the two-spin quantum model. We verified that, at the Ising limit ($\Delta\rightarrow\infty$), a similar method can also 
approximately describe the observed behavior .\par
\begin{figure}
 \centering
\includegraphics[scale=0.55]{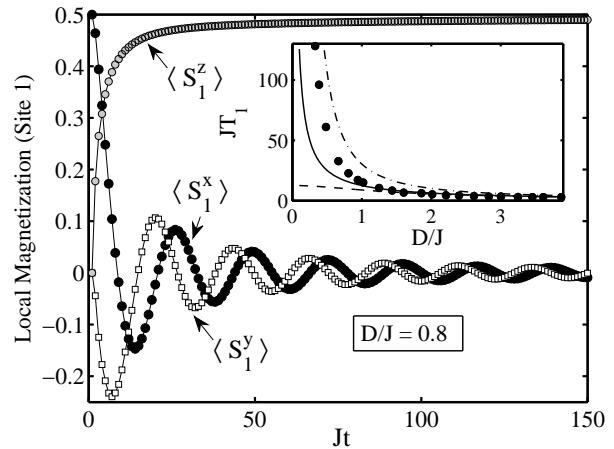}
\caption{Time evolution of local magnetization for the first spin along the chain for the arbitrary value $D/J=0.8$. Inset: Period of oscillations of 
the $x$-component of local magnetization as a function of $D/J$. The solid line is the curve $T_R=2\pi/\omega_R$, the dashed one is the curve 
$T_u=2\pi/\omega_u$, the dash-dotted line is the curve $T_h=2\pi/(\omega_u-\omega_d)$, and the symbols are the results from t-DMRG.\label{RabiP}}
\end{figure}
In Fig. \ref{Szj-JD} we show a snapshot of the local magnetization and spin-spin correlations in the $x$-direction at the arbitrary time $Jt=70$ and 
$D/J=0.8$. As can be seen, inhomogeneous spin-spirals with opposite chiralities propagate again from the boundaries towards the center of the chain. 
As shown in the classical picture projected on the $xz$-plane when $Jt=2$, the velocity of propagation is, in general, greater than in the case with 
$J=0$. As expected, the presence of EC promotes the onset of strong long-ranged spin-spin correlations (see Fig. \ref{Szj-JD}(c)). However, these 
correlations seems to be large for spin pairs (with respect to the center of the chain) that have not been touched by the spiral perturbations and 
diminish strikingly to zero for those that have undergone single-site oscillatory dynamics for a considerable time; that is, the propagating 
spin-spirals with opposite chiralities are almost uncorrelated. We also observe from Fig. \ref{Szj-JD}(b) that the nearest-neighbor spin-spin 
correlations are strongly weakened compared to the case with $J=0$, going to negligible values at (and near) the chain boundaries. A remarkable feature 
from Fig. \ref{Szj-JD}(a) is that in the presence of EC, the $y$-component of local magnetization is no longer ``quenched'', as it was in the case 
with $J=0$. A closer look at the single-site quantum dynamics is shown in Fig. \ref{RabiP}, where the time evolution of local magnetization for the 
first site along the chain is calculated. We immediately see that beginning in the state $\mid\rightarrow\rangle$, this spin precesses (or ``spirals'') 
clockwise towards the asymptotic state $\mid\uparrow\rangle$ with a frequency which we shall investigate next. We verified that the decay of 
oscillations is exponential. At the other boundary, the spin precesses counterclockwise towards the asymptotic state $\mid\downarrow\rangle$ with the 
same frequency. These perturbations are transmitted to the spins in the bulk apparently in a diffusive way, in contrast to the ballistic propagation 
which is observed in the case with $J=0$.\par
Based on the experience of the former section, we may want to consider the two-spin quantum dynamics of the model with DMI and isotropic EC. In Fig. 
\ref{Elev}(b) we show the energy diagram with $\omega_d=J/2$ and $\omega_u=\sqrt{\omega_R^2+\omega_d^2}$. In this case, the energy gap between the 
ground and first excitated state (which remains two-fold degenerate) is larger than in the case with $J=0$. The $x$-component of local magnetization of 
the spin par oscillate in phase 
\begin{equation}\label{Sx1}
 2\langle S_1^x\rangle=\cos(\omega_dt)\cos(\omega_ut)+\dfrac{\omega_d}{\omega_u}\sin(\omega_dt)\sin(\omega_ut).
\end{equation}
The nature of these oscillations depends strongly on the ratio $\omega_d/\omega_u=\sqrt{1+(D/J)^2}^{\;-1}$ which coincides with the anisotropy 
$J^{z}/J^{xx}$ in Eq.  \eqref{XXZ}. Then, when $D/J\gg1$ the leading behavior is governed by the first term of the sum in Eq. \eqref{Sx1}. This term 
describes fast oscillations with period $T_u=2\pi/\omega_u$ modulated by the cosine curve $\cos(\omega_dt)$, i.e, a beat-like pattern. According to 
Eq. \eqref{XXZ}, in this regime, the many-body system tends to be described by the XX model and we expect that these fast oscillations will resemble the 
Rabi oscillations found in the former section. To test these predictions, we calculate the period of oscillations (by averaging the time between 
sequential troughs) for the first spin along the chain as a function of $D/J$, using t-DMRG. This is shown in the inset of Fig. \ref{RabiP}. There, we 
see how the dashed line (representing $T_u$) tends to coincide with the solid one (representing $T_R=2\pi/\omega_R$) in the regime $D/J\gg1$, and both 
curves well describe the actual period of the single-site oscillations in the many-body system. On the other hand, when $0<D/J\ll1$, the behavior in 
Eq. \eqref{Sx1} is approximately well described by the curve $\cos[(\omega_u-\omega_d)t]$ which displays oscillations with the long period 
$T_h=2\pi/(\omega_u-\omega_d)$. This is represented in the inset of Fig. \ref{RabiP} by the dash-dotted line. At the limit $D/J\rightarrow0$ (isotropic 
Heisenberg model), the spin oscillations are not expected to be observed, as inferred by the divergence of the period of oscillations. In fact, our 
initial spin-polarized state in the $x$-direction is a stationary state of this model and hence no time evolution is detected in local observables, as 
we have verified numerically. It is curious to note that the period of spin oscillations in the two regimes can be remembered easily if we make 
an analogy with a radiative electronic transition. It is well known from elementary quantum mechanics that if an electron is undergoing a transition 
between stationary states $n$ and $l$, its position oscillates sinusoidally with an angular frequency $(E_n-E_l)/\hbar$. Thus, in the regime $D/J\gg1$, 
we might think of the spins as ``undergoing'' radiative transitions between the ground and first excited state (see Fig. \ref{Elev}(b)) and hence 
oscillating with the period $T_{<}\approx2\pi/(\omega_u+\omega_d)$, and in the regime $0<D/J\ll1$, they might be thought of as undergoing radiative 
transitions between the first and highest excited state and hence oscillating with the period $T_{>}\approx2\pi/(\omega_u-\omega_d)$.\par
One may ask about the validity of using the quantum two-spin model to interpret results from the many-body system in the present case. In the former 
section we argued that in the absence of EC this can be done for a certain time interval, since long-ranged spin-spin correlations were almost ``turned 
off'' during the spiral propagation and only the transport of nearest-neighbor spin-spin correlations were responsible for the \emph{onset} of quantum 
many-body effects, which make the predictions from the quantum two-spin model fail. In the presence of EC, we see that strong long-ranged spin-spin 
correlations develop in the system, and yet the single-site spin oscillations can be described approximately well by the quantum two-spin model. We 
believe that this happens due to the fact that only nearest-neighbor spin-spin correlations carry the information of relevant many-body effects and 
these correlations are very small in the presence of EC, especially at the boundaries (see Fig. \ref{Szj-JD}(b)). As noted from the inset of Fig. 
\ref{RabiP}, this picture is very suggestive, since there are no ``emergent'' Rabi frequencies in the presence of EC; i.e., the frequencies predicted 
by the quantum two-spin model are those approximately observed in the single-site dynamics of the many-body system. This was not the case in the 
absence of EC, since nearest-neighbor spin-spin correlations were remarkably larger during the spiral propagation. 
\begin{figure}
 \centering
\includegraphics[scale=0.55]{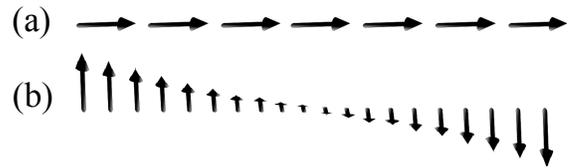}
\caption{(a) Initial spin-polarized state in the $x$-direction. (b) Approximate classical picture of the appearance of the state after the spin-spirals 
with opposite chiralities interfere at the center of the chain (see the text for deviations).\label{SDW}}
\end{figure}
\section{Comments on the long-time dynamics}\label{secLT}
In the previous sections, we studied the short-time dynamics of the spin-spiral propagation and oscillations in a spin chain with DMI and with(out) EC. 
By this, we always meant the time evolution just after the quench and \emph{before} the two perturbations coming from the boundaries interfere at the 
center of the chain. Even though we are not going to go deeper into what happens after this, it is worthwhile saying something for the sake of 
completeness. As might be expected intuitively, when the spin-spirals with opposite chiralities encounter one another, complex interference patterns arise. 
We observed in all cases that the components of local magnetization transverse to the direction of the DMI take on very small values as each propagating 
perturbation goes across the opposite half of the chain from where it was created, which is reasonable, due to the ``destructive'' character of 
interference in the $xy$-plane: spins in different halves of the chain have opposite direction of rotation (when they precess) in the $xy$-plane before 
the propagation, and almost uncorrelated spin-spirals reach the center. Then, beginning with our initial spin-polarized state in the $x$-direction as 
depicted in Fig. \ref{SDW}(a), we arrive after a long time at a state with an appearance approximately similar to that shown classically in Fig. \ref{SDW}(b), 
but with small fluctuations in the $xy$-plane which support the propagation of residual spin waves. Since we could not access very long times with our 
numerical calculations, it was not possible to study properties in the very long-time limit.
\section{Conclusions}\label{secC}
We have investigated the non-equilibrium dynamics of a spin-$1/2$ chain with Dzyaloshinskii-Moriya interactions in the presence (and absence) of 
Heisenberg exchange interactions after a quench by a transverse field. We found in all cases that inhomogeneous spin-spirals with opposite chiralities 
propagate from the boundaries towards the center of the chain until they meet at the center, producing complex interference patterns. This propagation 
is accompanied by quantum spin oscillations that decay asymptotically over time. We showed that important information for the interpretation of the 
observed quantum behavior can be extracted from classical considerations of the spin system, as well as from the quantum two-spin model of interactions. 
Our results contribute to recent efforts to understand the characteristics of the non-equilibrium dynamics following a quantum quench in closed 
one-dimensional spin systems, and are expected to be helpful for possible potential applications where the Dzyaloshinskii-Moriya interaction plays a 
relevant role in low-dimensional systems. 
\section*{acknowledgments}
 The author E.S.C. thanks the support from the Fulbright-Colciencias fellowship and the Mazda Foundation for the Arts and the Science fellowship. The 
support from the Direcci\'on de Investigaciones Sede Bogot\'a of the Universidad Nacional de Colombia is also acknowledged.

\bibliography{references}

\end{document}